\documentstyle[12pt]{article}

\large
\newcommand{\be}{\begin{eqnarray}}
\newcommand{\ee}{\end{eqnarray}}
\newcommand\del{\partial}

\begin{document}
\setlength{\baselineskip}{21pt}
\pagestyle{empty}
\vfill
\eject
\begin{flushright}
SUNY-NTG-93/40
\end{flushright}

\vskip 2.0cm
\centerline{\bf Chiral random matrix theory and the spectrum of}
\vskip 0.2 cm
\centerline{\bf the Dirac operator zero virtuality}
\vskip 2.0 cm
\centerline{Jacobus Verbaarschot}
\vskip .2cm
\centerline{Department of Physics}
\centerline{SUNY, Stony Brook, New York 11794}
\vskip 2cm

\centerline{\bf Abstract}
We study the spectrum of the QCD Dirac operator near zero virtuality.
We argue that it can be described by a random matrix theory
with the chiral structure of QCD.
In the large $N$ limit, this model
reduces to the low energy limit of the QCD partition function
put forward by Leutwyler and Smilga. We conjecture that the microscopic
limit of its spectral density is universal and reproduces that of QCD.
Using random matrix methods we obtain its exact analytical expression.
This result is compared to numerically calculated spectra for a
liquid of instantons, and we find a very satisfactory agreement.

\vfill
\noindent
\begin{flushleft}
SUNY-NTG-93/40\\
October 1993
\end{flushleft}
\eject
\pagestyle{plain}

\vskip 1.5cm
\renewcommand{\theequation}{1.\arabic{equation}}
\setcounter{equation}{0}
\noindent
\centerline{\bf 1. Introduction}
As has become clear from a wide body of both numerical and theoretical
evidence, the $SU(N_f)\times SU(N_f)$ chiral symmetry of the massless QCD
Lagrangian is broken spontaneously to $SU(N_f)$, whereas the $U_A(1)$
symmetry is broken explicitly by the anomaly (see \cite{THOOFT-1986}).
The order parameter
that characterizes the presence of chiral symmetry is the chiral
condensate. According to the Banks-Casher formula \cite{BANKS-CASHER-1980},
it is proportional
to the spectral density of the Euclidean Dirac operator
near zero virtuality. Therefore, the behaviour of
the spectrum in this region is of fundamental importance to the phenomenon
of chiral symmetry breaking.

In this paper we study the spectrum of the QCD Dirac operator
near zero virtuality in the thermodynamic
limit while keeping the average spacing between the eigenvalues
fixed, $i.e.$ the spectrum is blown up by a factor $V_4$ as
$V_4\rightarrow\infty$. This
limit of the spectral density,
which has a nontrivial limit as a consequence of a finite value of the
chiral condensate, will be called the microscopic spectral density.
It measures correlations on the order of one level spacing as opposed to
the average spectral density that is obtained by averaging over many levels.
We conjecture that the microscopic spectral density is $universal$.
This is reminiscent to the universality
observed in the theory of critical phenomena where large scale fluctuations
wash out the details of the interaction.  Because
the spectrum near zero virtuality is dominated by the soft modes
of the system, we expect that it is entirely determined by the
symmetries of the microscopic theory.
Indeed, Leutwyler and Smilga \cite{LEUTWYLER-SMILGA-1993}
were able to express negative
moments of the microscopic spectral density in terms of the effective
low-energy limit of QCD that only embodies its underlying chiral structure.

A similar splitting of scales between spectral fluctuations on the
order of one level spacing and the average spectral density is well-known
in the statistical theory of spectra \cite{BRODY-ETAL-1981} and quantum
chaos \cite{BOHIGAS-GIANNONI-1984,SELIGMAN-VERBAARSCHOT-ZIRNBAUER-1984}.
In that case one finds that correlations
on the microscopic scale are universal if the classical counterpart of
the system is chaotic. The correlations only depend on the symmetries
of the system and are given by the invariant random matrix ensembles.
This suggests to describe the microscopic spectral density of the
QCD Dirac operator in terms of a random matrix theory. The relevant
ensemble will be called the chiral random matrix ensemble. We have
shown \cite{SHURYAK-VERBAARSCHOT-1993}
that it is equivalent to the low-energy limit of the chiral
effective Lagrangian (see section 3). Moreover, techniques developed
in random matrix theory enabled us \cite{VERBAARSCHOT-ZAHED-1993}
to obtain an analytical expression for the microscopic
spectral density (see section 4).

Apart from the above arguments, there is additional evidence in favor of the
universality of the spectrum near zero virtuality. First,
the Leutwyler-Smilga sum rules are not only found in QCD but
in the Schwinger model as well \cite{SMILGA-1993} and, numerically,
for an instanton liquid model
of the QCD-vacuum \cite{SHURYAK-VERBAARSCHOT-1993}.
Second, the chiral random matrix model reproduces all
Leutwyler-Smilga sum rules \cite{SHURYAK-VERBAARSCHOT-1993}.
Third,  the microscopic spectral density  of a Dirac operator in
a liquid of instantons agrees within numerical accuracy
with  the random matrix result (see section 6). Fourth, the same microscopic
spectral density is found in the Hofstadter model for universal conductance
fluctuations \cite{SLEVIN-NAGAO-1993}.

The Leutwyler-Smilga sum rules and the microscopic spectral density are
based on the observation that a nonzero value of the chiral condensate implies
that the spacing of the
small eigenvalues is $\sim 1/V_4$, as opposed to $\sim 1/V_4^{1/4}$
for a noninteracting system. Such accumulation of small eigenvalues is obtained
naturally for a liquid of instantons, where the exact fermionic zero modes
can be chosen as basis states to construct a matrix representation
of the Dirac operator \cite{DIAKONOV-PETROV-1985B}. Such model can be simulated
numerically and results for its spectral density are presented in section 7.

This paper is organized as follows. In section 2 we give an outline of the
general framework of this paper and
introduce the microscopic
spectral density. A derivation of the simplest Leutwyler-Smilga sum rule
is given in section 3. In section 4 we formulate the chiral random
matrix model for the trivial topological sector of the QCD vacuum. Its
equivalence to the low-energy effective partition function is shown in
section 5. The microscopic spectral density is calculated in section 6, and,
in section 7, numerical results for a liquid of instantons are presented.
Concluding remarks are made in section 8.

\vskip 1.5cm
\renewcommand{\theequation}{2.\arabic{equation}}
\setcounter{equation}{0}
\centerline{\bf 2. Formulation of the problem}
\vskip 0.5 cm
In this paper we study the spectrum of the massless Euclidean Dirac operator
defined by the eigenvalue equation

\be
(i\gamma \del + \gamma A) \phi_\lambda = \lambda \phi_\lambda,
\ee
where $A$ is an $SU(N_c)$ valued gauge field. The statistical distribution
of the eigenvalues
is induced through the gauge field configuration by the statistical weight
of the Euclidean QCD partition function
\be
Z = \sum_\nu e^{i\nu\theta} < \prod^{N_f}_{f=1}\prod_{\lambda_n>0}
(\lambda_n^2 + m_f^2) m_f^{\nu}>_{S_\nu(A)},
\ee
where the masses play the role of sources.
The product is over the positive eigenvalues of the Dirac operator,
and the average $<\cdots>_{S_\nu(A)}$ is over gauge field configurations
with topological quantum number $\nu$ weighted by the gauge field action
$S_\nu(A)$. The topological part of the action, $\exp i\theta\nu$,
has been displayed explicitly.
The eigenvalues occur in pairs, $\lambda$ and $-\lambda$, with corresponding
eigenfunctions $\phi_\lambda$ and $\gamma_5\phi_\lambda$. The only exception
is $\lambda = 0$, when we have the logical possibility that
$\gamma_5 \phi_\lambda \sim \phi_\lambda$. Indeed, according to the
Atiyah-Singer theorem, gauge field configurations with topological quantum
number $\nu$ have at least $\nu$ zero modes with a definite chirality.

The propagator is given by
\be
S(x,y) = \sum_\lambda -\frac {\phi_\lambda(x) \phi_\lambda^*(y)}{\lambda+im},
\ee
where we have suppressed the color and Dirac indices.
The chiral condensate $<\bar q q>$
is defined as the space-time average of $<{\rm Tr}S(x,x)>$ (the brackets
$<\cdots>$ without indices denote averaging with respect to the
partition function under consideration).
According to the Banks-Casher \cite{BANKS-CASHER-1980} formula we have
\be
<\bar q q> = \frac 1{V_4} \int d^4 x <{\rm Tr} S(x,x)> = i \pi \frac{<\rho(0)>}
{V_4},
\ee
where the spectral density $\rho(\lambda)$ is defined as
\be
\rho(\lambda) = \sum_{\lambda_n} \delta(\lambda -\lambda_n).
\ee
The condensate can be expressed as a derivative of the partition function
\be
<\bar q q> = \lim_{m_f\rightarrow 0} \lim_{V\rightarrow \infty}  \frac i{V_4}
\frac d{dm_f} \log Z(m_f).
\ee
The spectral density at zero virtuality has to be defined with care: the
thermodynamic limit has to be taken before the chiral limit, $i.e.$,
\be
\rho(0) = \lim_{m\rightarrow 0} \lim_{V_4\rightarrow \infty} \frac 1\pi
\sum_\lambda \frac m{\lambda^2 +m^2}.
\ee
In this way it is possible to obtain a nonzero spectral density $<\rho(0)>$
even in the case that  zero eigenvalues are absent at all $finite$ volumes.
If this leads to a nonzero value of the chiral condensate, this phenomenon is
referred to as the spontaneous breaking of chiral symmetry. A necessary
requirement for this to happen is that
\be
<\rho(0)> \sim V_4,
\ee
or, put differently, the spacing between the eigenvalues near zero virtuality
is $\sim 1/V_4$.

This number should be contrasted with the spacing of the small eigenvalues
for a non-interacting system which is $\sim 1/V^{1/4}$ (obtained from the
solution of the free Dirac equation in a hypercubic box of volume $V_4$).
A natural explanation for the accumulation of eigenvalues near zero
is provided by the properties of gauge field configurations
of a liquid of instantons.
Each isolated instanton, a localized
field configuration by itself,  gives rise
to one localized zero mode. When the space-time volume is saturated with
an equal number of instantons and anti-instantons the low-lying modes can be
expressed as linear combinations of the exact zero modes and form
the zero mode zone \cite{SHURYAK-1988}.
For a rather dilute ensemble we therefore expect that
the number of small eigenvalues is $\sim V_4$.

As was observed by Leutwyler and Smilga, eq. (2.4) implies the existence
of a family of new sum rules. The simplest one involves the sum
\be
\frac 1{V^2_4} \sum_{\lambda_n\ne 0} <\frac 1{\lambda^2_n}>_\nu
\ee
which should converge to a finite limit for $V_4 \rightarrow \infty$.

The most important
consequence of chiral symmetry breaking is the appearance of a massless
pion in the chiral limit ($m \rightarrow 0$). In other words, the pion
correlator at zero momentum (note that $S(y,x) = -\gamma_5 S^\dagger(y,x)
\gamma_5$)
\be
\Pi(p=0) = \int d^4y < {\rm Tr} S(x,y) S^\dagger(x,y) >
\ee
diverges as $\sim 1/m$. This follows from the chiral
Ward identity
\be
\Pi(p=0) = \frac 1{2im} (<{\rm Tr} S(x,x)> - <{\rm Tr}
S^\dagger(x,x)>),
\ee
which allows us to express this correlator
in the spectral density:
\be
\Pi(p=0) = \frac 1{mV_4} \pi<\rho(0)>,
\ee
where we have used the translational invariance of $<{\rm Tr} S^\dagger(x,x)>$.
{}From eq. (2.4) it then immediately follows that a finite value of the
condensate leads to a massless pion in the chiral limit. As always,
it is understood that the thermodynamic limit is to be taken before
the limit $m \rightarrow 0$.

Because of the importance of the small eigenvalues in the spontaneous breaking
of chiral symmetry, we propose to study
the spectrum near zero virtuality in the thermodynamic by enlarging it
proportional to $V_4$. In this way the spacing of the eigenvalues remains
${\cal O}(1)$ for $V_4\rightarrow\infty$.
We call this limit the microscopic limit. Formally, it is defined as
\be
\rho_S(x) = \lim_{ V_4\rightarrow \infty} \frac 1V_4 <\rho(\frac xV_4)>_\nu
\ee
in the sector of topological charge $\nu$.
All Leutwyler-Smilga sum rules can be expressed in the microscopic spectral
density and
similarly defined spectral correlation functions. For the
sum (2.9) we find
\be
\int dx \frac{ \rho_S(x)}{x^2}.
\ee
By  using random matrix theory we will obtain
an analytical expression for
$\rho_S(x)$ in the zero topological charge sector.

\vskip 1.5cm
\renewcommand{\theequation}{3.\arabic{equation}}
\setcounter{equation}{0}
\centerline{\bf 3. Derivation of Leutwyler-Smilga sum rules}
\vskip 0.5 cm

According to Leutwyler and Smilga \cite{LEUTWYLER-SMILGA-1993},
the mass dependence of the low-energy limit of the QCD partition
function is given by
\be
\frac{Z(m,\theta)}{Z(m=0,\theta)}= \int_{\det U = 1} {\cal D}U
\exp(\frac{V_4\Sigma}{2}{\rm Tr}(m U^{-1} \exp(-i\theta/N_f)
+m U \exp(i\theta/N_f)).
\ee
The integration is over $SU(N_f)$ with the Haar measure, $m$ is the quark
mass matrix, which is taken diagonal,
and $\Sigma$ is the the absolute value of chiral condensate.
For one flavor, when the integration is absent, this ratio simplifies to
\be
\frac{Z(m,\theta)}{Z(m=0,\theta)} = \exp(mV_4\Sigma \cos\theta).
\ee
Expanding this ratio in powers of $m$ and comparing the coefficients to those
of the same expansion of the full QCD partition function provides us
with an infinite family of sum rules. The simplest sum rule is derived
from the ${\cal O}(m^2)$ term. In the zero topological
charge sector (projected out by integrating over $\theta$) the result
is
\be
\sum_{\lambda > 0} <\frac 1{\lambda^2 V^2_4}>_{\nu = 0} =
\frac{\Sigma^2}{4 N_f},
\ee
where the case $N_f=1$ follows immediately from eq. (3.2).

\vskip 1.5cm
\renewcommand{\theequation}{4.\arabic{equation}}
\setcounter{equation}{0}
\centerline{\bf 4. The chiral random matrix model}
\vskip 0.5 cm

In this section we construct a random matrix model for the spectrum of the
Dirac operator near zero virtuality in the sector of zero total topological
charge. The underlying assumption is that
the microscopic spectral density is universal, and only follows from the
symmetries of the system.
For $N_f$ flavors, the partition function
that reflects the chiral structure of QCD is given by
\cite{NOWAK-VERBAARSCHOT-ZAHED-1989A,SIMONOV-1991,SHURYAK-VERBAARSCHOT-1993}
\be
Z = \int {\cal D}T P(T)\prod_f^{N_f}\det \left (
\begin{array}{cc} m_f & iT\\
                 iT^\dagger & m_f
\end{array} \right ),
\ee
where the integral is over the real and imaginary parts of the matrix elements
of the arbitrary complex  $N/2\times N/2$ matrix $T$,
$i.e.$, ${\cal D}T$ is the Haar measure.
In agreement with the maximum entropy principle \cite{BALIAN-1968}
the distribution function of the overlap matrix
elements $P(T)$ is chosen Gaussian
\be
P(T) = \exp(-\frac {N\Sigma^2}{2} {\rm Tr }T T^{\dagger}).
\ee
The symplectic structure is a manifest consequence of chiral symmetry,
and implies that the quark eigenvalues occur in pairs. The density of the
modes that define the matrix representation of the Dirac operator, $N/V_4$,
is taken equal to 1, which allows us to identify
the space-time volume and the total number of modes $N$.

The partition function (4.1) is inspired by the instanton liquid approximation
to the QCD-vacuum \cite{DIAKONOV-PETROV-1986,SHURYAK-1988}, where the fermion
determinant is approximated by its value in the space of the fermionic zero
modes. In that case, the overlap matrix elements $T_{ij}$ are a function of the
collective coordinates of the instantons. Therefore, they are not statistically
independent, and, as we will see in section 7, the average level density
differs from a semi-circle, the result that can be
obtained for the ensemble (4.1).

Although in this section the random matrix model has been defined
for zero topological charge only, it can be easily extended
\cite{SHURYAK-VERBAARSCHOT-1993} to the general
case including an arbitrary value of the QCD $\theta-$angle. The essential
observation is that if $T$ is an $N\times M$ rectangular matrix, the matrix
in eq. (4.1) has exactly $|N-M|$ zero eigenvalues for $m_f = 0$.
For simplicity we restrict ourselves in this paper to square overlap
matrices.

\vskip 1.5cm
\renewcommand{\theequation}{5.\arabic{equation}}
\setcounter{equation}{0}
\centerline{\bf 5. Reduction of the partition function}
\vskip 0.5 cm
In order to evaluate the partition function (4.1)
the determinant is written as an integral over Grassmann
variables
\be
\prod_f\det \left ( \begin{array}{cc} m_f &i T
\\ iT^\dagger & m_f \end{array}\right)
= \int \prod_f {\cal D}\psi^f {\cal D}\phi^f \exp\sum_f
\left ( \begin{array}{c} \psi^{f*} \\ \phi^{f*} \end{array}\right)
\left ( \begin{array}{cc} m_f & iT \\ iT^\dagger & m_f \end{array}\right)
\left ( \begin{array}{c} \psi^f \\ \phi^f \end{array}\right),\nonumber\\
\ee
where the measure of the Grassmann integration is as usual
\be
{\cal D} \psi^f = \prod_i d\psi_i^f d\psi_i^{f*},
\ee
and the conjugation $^*$ is the  conjugation of the second kind
($i.e.$, $\psi^{**} = -\psi$)
\cite{EFETOV-1983,VERBAARSCHOT-WEIDENMUELLER-ZIRNBAUER-1985}.
The integral over $T$ is gaussian and can
be performed easily. In the partition function this results
in the factor
\be
\exp \frac{2}{N\Sigma^2}
\psi^{f*}_i \psi_i^g \phi_j^{g*}\phi_j^f,
\ee
which represents a 4-fermion interaction.

The quartic term can be written as a sum of two squares
\be
\psi^{f*}_i \psi_i^g \phi_j^{g*}\phi_j^f =
\frac 14( \psi^{f*}_i \psi_i^g +  \phi_i^{f*}\phi_i^g)
(\psi^{g*}_j \psi_j^f + \phi_j^{g*}\phi_j^f )
-\frac 14( \psi^{f*}_i \psi_i^g -  \phi_i^{f*}\phi_i^g)
(\psi^{g*}_j \psi_j^f- \phi_j^{g*}\phi_j^f ).\nonumber\\
\ee
Each of the two squares can be linearized with the help of
a Hubbard-Stratonovich transformation
\cite{VERBAARSCHOT-WEIDENMUELLER-ZIRNBAUER-1985}. This allows us to perform
the Grassmann integrations at the expense of the introduction of the new
real valued integration variables $\sigma^{fg}$ and $\bar\sigma^{fg}$,
respectively. Apart from an irrelevant overall constant, the
partition function reduces to
\be
Z = \int {\cal D}\sigma {\cal D}\bar\sigma{\det}^{N/2}(\sigma +i\bar\sigma
+m) {\det}^{N/2}(\sigma -i\bar\sigma+m)
\exp-\frac{N\Sigma^2}{2}{\rm Tr} (\sigma+i\bar\sigma)(\sigma-i\bar\sigma).
\ee
As always, the measure of the integral over the
matrices $\sigma$ and $\bar \sigma$ is the
Haar measure. The diagonal mass matrix is denoted by $m$.
This integral greatly simplifies for $N_f = 1$. This case is discussed
in Appendix A where we obtain Leutwyler-Smilga sum rules for
finite values of $N$.

The complex matrix $\sigma + i\bar\sigma$ can be decomposed in 'polar
coordinates' as \cite{HUA-1963}
\be
\sigma + i\bar\sigma = U \Lambda V^{-1},
\ee
where $U$ and $V$ are unitary matrices and $\Lambda$ is a diagonal real
positive definite matrix. Since the r.h.s has $N_f$ more degrees of freedom
than the l.h.s., one has to impose constraints on the new integration
variables.
This can be achieved \cite{HUA-1963}
by restricting $U$ to the coset $U(N_f)/U(1)^{N_f}$, where
$U(1)^{N_f}$ is the diagonal subgroup of $U(N_f)$.
In terms of the new variables the partition function reads
\be
Z &=& \int J(\Lambda){\cal D}\Lambda {\cal D}U{\cal D} V
{\det}^{N/2}(U\Lambda V^{-1}+m)
{\det}^{N/2}(V\Lambda U^{-1}+m)\exp(-\frac{N\Sigma^2}{2}{\rm Tr} \Lambda^2),
\nonumber\\
\ee
where the integral over $U$ is over $U(N_f)/U(1)^{N_f}$ and the integral
over $V$ is over $U(N_f)$.

For $N_f$ flavors we have $N_f$ condensates which break down the symmetry of
the action to $U(N_f)/U(1)^{N_f}$ leaving us with $N_f^2$ Goldstone
modes. When we allow for fluctuations of the topological charge
the phase of the determinant becomes massive and the usual number of
of $N_f^2 -1$ Goldstone modes is recovered.

The term proportional to $m$ plays the role of a small symmetry breaking
term. The integrals over the nonzero modes will
be performed by a saddle point integration at $m = 0$,
whereas the integrals over the
soft modes will be accounted for exactly for a fixed value of $mN$.
The nonzero mode part of the partition function is given by
\be
Z(m=0) = \int J(\Lambda) d\Lambda
{\det}^N\Lambda \exp(-\frac{N\Sigma^2}{2}{\rm Tr} \Lambda^2).\nonumber\\
\ee
The leading order contribution in $1/N$ of the integral over $\Lambda$
can be obtained by a saddle point
approximation. The saddle point equations for the $\Lambda$ integrals
read
\be
\Lambda_i = \pm \lambda.
\ee
The negative solution is not inside the integration manifold and can be
omitted.

At the saddle point in $\Lambda$,
the $U-$dependence can be absorbed into $V$. The
the $U-$integration yields a finite irrelevant constant.
The remaining integral over $V$ which is over $U(N_f)$ can be split
into an integral over the phase of the determinant and an integral
over $SU(N_f)$.
We treat $m$ as a small parameter and expand the
determinants up to first order in $m$. The result for the $m$ dependent
part of the partition function is
\be
\frac{Z(m)}{Z(m=0)}= \int_0^{2\pi} d\alpha
\int_{\det V = 1} {\cal D}V
\exp(\frac{N\Sigma}{2}{\rm Tr}(m V^{-1} \exp(-i\alpha/N_f)
+m V \exp(i\alpha/N_f)),\nonumber\\
\ee
which coincides with the result for zero topological charge
derived by Leutwyler and Smilga
\cite{LEUTWYLER-SMILGA-1993} for the QCD partition function using chiral
perturbation theory. As was shown in \cite{SHURYAK-VERBAARSCHOT-1993}
this result can be easily generalized to an arbitrary topological charge
sector.

The value of the chiral condensate can be obtained from eq. (2.6). In the
case of equal positive masses we have
\be
<\bar q q>_{\nu=0} = \lim_{m\rightarrow 0}
\lim_{N\rightarrow \infty} \frac \Sigma{2N_f}
<{\rm Tr} (V+ V^{-1})>_{\nu = 0}.
\ee
For $N_f = 1$ the integral over $V$ is absent, and
the sign of the quark condensate
is independent of the sign of $m$. For more than one flavor the order of the
limits allows us to perform
the $V$ integral by a saddle point approximation.
In the case of equal positive masses the saddle point is
at $V = {\bf 1}$ which allows us to identify the parameter $\Sigma$
and the chiral condensate
\be
<\bar q q> = \Sigma.
\ee
This completes the reduction of the partition function.

As observed in \cite{LEUTWYLER-SMILGA-1993},
for more than one flavor the value of the condensate depends on the sign of
the quark mass. For example, in the case of two flavors with equal negative
masses and $\theta = 0$, the saddle point is at $V = {\bf -1}$.

\vskip 1.5cm
\renewcommand{\theequation}{6.\arabic{equation}}
\setcounter{equation}{0}
\centerline{\bf 6. Spectral density of the random matrix model}
\vskip 0.5 cm
In this section we obtain the spectral density of the random matrix partition
function (4.1). We use the orthogonal polynomial method
developed by Wigner, Dyson and Mehta \cite{MEHTA-1991,PORTER-1965} in the
context of the invariant random matrix ensembles. The first step is
to rewrite the matrix integration in
the partition function (4.1) in polar coordinates.
For an arbitrary complex matrix this amounts to the transformation
\cite{HUA-1963}
\be
T = U \Lambda V^{-1},
\ee
where $U$ and $V$ are unitary matrices and $\Lambda$ is a positive definite
diagonal matrix. Since the r.h.s has $n\equiv N/2$ more degrees of freedom
than the l.h.s., one has to impose constraints on the new integration
variables.
This can be achieved \cite{HUA-1963}
by restricting $U$ to the coset $U(n)/U(1)^{n}$, where
$U(1)^{n}$ is the diagonal subgroup of $U(n)$. The Jacobian of this
transformation, that  depends only on the eigenvalues $\lambda_k$ of $\Lambda$,
is given by
\be
J(\Lambda) = \prod_{k < l}(\lambda^2_k -\lambda^2_l)^2 \prod_k \lambda_k.
\ee
The integrations over the eigenvalues and the unitary matrices decouple. The
latter only result in an overall irrelevant constant factor and can be ignored.
The eigenvalue distribution is thus given by
\be
\rho_n(\lambda_1, \cdots, \lambda_n) = J(\Lambda)
\prod_{f}\prod_k(\lambda_k^2 +m_f^2) \exp(-n\Sigma^2\sum_{k=1}^n
\lambda_k^2).
\ee
The spectral density $\rho_1(\lambda)$ is obtained by integration over
the remaining $n-1$ eigenvalues
\be
\rho_1 (\lambda_1) = \int \prod_{k=2}^n d\lambda_k
\rho_n(\lambda_1, \cdots, \lambda_n).
\ee
These integrals can be evaluated with the help of the orthogonal polynomial
method (see \cite{MEHTA-1991} for references).
The main ingredient is to write the product over the differences of
the eigenvalues as a Vandermonde determinant, $i.e.$
\be
\prod_{k < l}(\lambda^2_k -\lambda^2_l)^2 =
    \left | \begin{array}{ccc} 1     & \cdots  &  1     \\
                               \cdot &         & \cdot  \\
                               \cdot &         & \cdot  \\
                              \lambda_1^{2(n-1)} &\cdots &\lambda_n^{2(n-1)}
    \end{array}    \right|^2.
\ee
which up to a constant can be rewritten in terms of orthogonal polynomials
$P_k$ as
\be
    \left | \begin{array}{ccc} P_0(\lambda_1^2) & \cdots  & P_0(\lambda_n^2)\\
                               \cdot &         & \cdot  \\
                               \cdot &         & \cdot  \\
                       P_{n-1}(\lambda_1^2) &\cdots  &P_{n-1}(\lambda_n^2)
    \end{array}    \right|.
\ee
The $P_k$ will be chosen orthogonal according to the weight function
\be
\int_0^\infty d(\lambda^2) (\lambda^2 + m^2)^{N_f} \exp(-n\Sigma^2
\lambda^2) P_k(\lambda^2) P_l(\lambda^2) = \delta_{kl}.
\ee
For $m = 0$ these
polynomials are well known,
\be
P_k(s) = \left ( n\Sigma^2
\frac{k!}{\Gamma( N_f +k +1)}\right )^{1/2} L_k^{N_f}(sn\Sigma^2),
\ee
where the $L_k^{N_f}$ are the generalized Laguerre polynomials.

The determinant can be expanded according its definition.
All integrals
can be performed immediately by orthogonality and, up to an overall constant,
 we are left with
\be
\rho_1(\lambda) = 2\Sigma \sqrt{n}\sum_{k=0}^{n-1}
\frac{k!}{\Gamma( N_f +k +1)} L_k^{N_f}(z)
L_k^{N_f}(z) z^{N_f +1/2} \exp( -z),
\ee
where $z$ is defined by
\be
z = n\lambda^2\Sigma^2.
\ee
The normalization constant has been chosen such that $\int d\lambda
\rho(\lambda) = N$ (remind that $N=2n$).
The sum can be evaluated exactly with the Christoffel-Darboux formula
(which can be found in any treatise on orthogonal polynomials),
resulting in
\be
\rho_1(\lambda) = \frac{2\Sigma \sqrt{n}\, n!}
{\Gamma( N_f +n)}\left ( L_{n-1}^{N_f}(z)L_{n-1}^{N_f+1}(z)-
L_{n}^{N_f}(z)L_{n-2}^{N_f+1}(z)\right )z^{N_f +1/2} \exp( -z),
\ee
which, constitutes the exact spectral density of the model (4.1).
The microscopic limit is obtained by taking $N \rightarrow \infty$ while
keeping $N\lambda = x$ fixed (remember that $n = N/2$).
This can be achieved from the asymptotic relation
\be
\lim_{n \rightarrow \infty} \frac 1{n^\alpha} L_n^\alpha(\frac xn) =
x^{-\frac {\alpha}{2}} J_\alpha(2 \sqrt x),
\ee
where $J_\alpha$ is the ordinary Bessel function of degree $\alpha$.
One notices that to leading order in $n$ the difference in eq. (6.11) cancels.
However, using recursion relations for the generalized Laguerre polynomials,
this cancellation can be achieved explicitly, and the asymptotic relation
can be applied to the next to leading order terms.
The result for the microscopic spectral density is
\be
\rho_S(x)  = \frac {\Sigma^2 x}{2} (J^2_{N_f}(\Sigma x) -J_{N_f+1}(\Sigma x)
             J_{N_f-1}(\Sigma x)).
\ee
{}From the asymptotic relation for the Bessel function
\be
J_\nu(z) \sim \left ( \frac 2{\pi z} \right )^{1/2} \cos(z - \frac{\pi}2 \nu-
\frac {\pi}4)\quad {\rm for} \quad z\rightarrow \infty
\ee
we find that
\be
\lim_{x\rightarrow\infty} \rho_S(x) =  \frac {\Sigma}{\pi}
\ee
which agrees with the Banks-Casher relation. As follows from the
small $x$ behaviour of
\be
\rho_S(x) \sim \frac {\Sigma}{N_f!(N_f+1)!}\left
( \frac {\Sigma x}{2} \right)^{2N_f +1}\quad {\rm for}\quad x\rightarrow 0,
\ee
small eigenvalues are strongly suppressed for an increasing number of flavors.

The formula (6.13) reproduces all diagonal sum rules of Leutwyler and Smilga,
$e.g.$, the sum
\be
\sum_n \frac 1{N^{2p} \lambda_n^{2p}}
\ee
can be converted into an integral over the microscopic variable $x =\lambda N$
resulting in
\be
\int_0^\infty \frac {\rho_S(x) dx}{x^{2p}} =
\left(\frac {\Sigma}{2}\right)^{2p}\,\frac{\Gamma (2p-1)\Gamma (N_f-p+1)}
{\Gamma (p)\Gamma (p+1) \Gamma (N_f + p )}
\ee
The above spectral density thus summarizes all sum rules and yields
new sum rules, e.g. sum rules for noninteger values of $p$.

It is possible to derive all higher order spectral correlation functions which
generate an infinite family of new spectral sum rules. An explicit expression
for the level correlation function is given in \cite{VERBAARSCHOT-ZAHED-1993},
and it generates sum rules obtained previously by Leutwyler and Smilga
\cite{LEUTWYLER-SMILGA-1993}.
\newpage
\vskip 1.5cm
\renewcommand{\theequation}{7.\arabic{equation}}
\setcounter{equation}{0}
\centerline{\bf 7. Spectral density for the instanton liquid}
\vskip 0.5 cm
The sum over field configurations in the
partition function (2.2) can be approximated semiclassically by
that of
a liquid of instantons. Instead of averaging over all gauge field
configurations, we average over the collective coordinates of the
instantons only, whereas 1-loop quantum fluctuations about the instantons
are included in the measure. The action in (2.2) is the instanton
action, which also includes the interaction between instantons.
We use the so called streamline \cite{YUNG-1988,VERBAARSCHOT-1991}
interaction supplemented by a core in order to stabilize the instanton
liquid. The fermion determinant is calculated in the space spanned by
the fermionic zero modes with overlap matrix elements that can be
derived from the streamline configuration
\cite{SHURYAK-VERBAARSCHOT-1992A}. More details on the above instanton
liquid model can be found in \cite{SHURYAK-VERBAARSCHOT-1993A}.

The numerical simulations were carried out for a liquid of 64 instantons
in a Euclidean space time volume of $(2.378)^3\times 4.756$ in units of
$\Lambda^{-4}_{QCD}$. Averages were obtained from 1,000 statistically
independent configurations for $N_f = 1,\,2,\,3$ and 100,000 for $N_f =0$.
Our main results are presented in Figs.  1 and 2. The number of flavors
is shown in the label of the figures.
In Fig.~1 we show the average
spectral density $n(\lambda)$
of the Dirac operator in the space of the zero modes (the normalization
is $\int_0^\infty n(\lambda) d\lambda = 1$, and $\lambda$ is in units of
$\Lambda_{QCD}$, so $n(\lambda)$ differs from $<\rho(\lambda)>$ by a
normalization factor).
Note that for $N_f = 2$ the thermodynamic limit of the
slope $n'(0)$ is zero, whereas for $N_f= 3$ it is negative. This agrees
with a recent result of Smilga and Stern \cite{SMILGA-STERN-1993} according
to which the flavor dependence of $n'(0) $ is given by
$\sim(N_f^2 - 4)/N_f$.

For $N_f =0,\, 1,\, 2$ we observe a clear separation of the microscopic scale
near zero virtuality and the overall average spectral density. The latter
is obtained by a fit with a smooth function ignoring the region near
zero virtuality. This allows us of obtain a finite value for $<\rho(0)>$.
The microscopic limit of the spectral density is shown in Fig. 2. The full line
is the spectral density of the the
unfolded spectrum $<\rho(\mu)>$ (see ref. \cite{BOHIGAS-GIANNONI-1984}
for an exact definition),
for which the average spacing between the eigenvalues
is equal to one. The dashed curve shows the asymptotic result
for the microscopic spectral density in the same units which agrees well with
the random matrix result. This is even more remarkable if one realizes that
the average spectral density is very different from the random matrix
result (a semi-circle). It should be stressed that the microscopic spectral
density contains no free parameters. For $N_f =3$ it is no longer possible
to extract the thermodynamic limit of $<\rho(0)>$ and an unambiguous comparison
with the microscopic spectral density cannot be made. The strong supression of
small eigenvalues is consistent with eq. (6.16).

We observe that the spectrum is very stiff. The peak occurs because the
smallest eigenvalue is always close to its average position. As is also
the case for the other invariant random matrix ensembles, the fluctuations
are much less than for a random sequence of eigenvalues, although
it looks like that the oscillations in the asymptotic spectral density
are not reproduced by the liquid of instantons. This may be
a finite size effect.

The Leutwyler-Smilga sum rules can be checked for a liquid of instantons.
The results completely agree with the the exact
results in \cite{LEUTWYLER-SMILGA-1993} (see \cite{SHURYAK-VERBAARSCHOT-1993}
for a detailed comparison).

\vskip 1.5cm
\renewcommand{\theequation}{8.\arabic{equation}}
\setcounter{equation}{0}
\centerline{\bf 8. Conclusions}
\vskip 0.5 cm

We have studied the spectrum of the
QCD Dirac operator near zero virtuality with the help
of the microscopic
spectral density, which is obtained by enlarging the spectrum by  $V_4$
as $V_4 \rightarrow \infty$. The existence of this limit is an
immediate consequence of the spontaneous breaking of chiral symmetry.

We have conjectured that the microscopic spectral density is universal. This
should be contrasted with the average spectral density which depends
on the details of the system. The latter varies on a scale of
${\cal O}(V^0_4)$, whereas the first shows fluctuations on the
microscopic scale of ${\cal O}(1/V_4)$. The implicit assumption is
that both scales can be separated. On the basis of the universality
we have formulated
a random matrix theory that apart form the chiral structure of
the QCD Dirac operator has no other information content.

It was shown that this random matrix theory is equivalent to
the extreme low-energy limit
of the QCD partition function. Therefore all Leutwyler-Smilga
sum rules are reproduced.
Using the orthogonal polynomial technique
developed by Wigner, Dyson and Mehta, we were able to obtain the exact
analytical result for the microscopic spectral density.

In general it is very hard to obtain exact numerical spectra for a field
theory. However, in the instanton liquid model for the QCD vacuum we were
able to obtain sufficient statistics for the microscopic
spectral density, and we found a very satisfactory agreement with the
random matrix model.
In view of this, it would be very interesting
to compare our results to spectra obtained
from lattice QCD.

\vskip 1.5cm
\renewcommand{\theequation}{A.\arabic{equation}}
\setcounter{equation}{0}
\noindent
\centerline{\bf Appendix A}
\vskip 0.5 cm
In this appendix we derive the sum rules for one flavor and finite values
of $N$. For $N_f = 1$ the partition
function (5.5) simplifies to
\be
Z = \int d\sigma d\bar\sigma (\sigma + i\bar\sigma -m)^{\frac N2}
(\sigma - i\bar\sigma -m)^{\frac N2}
\exp(-\frac{N\Sigma^2}{2}(\sigma+ i\bar\sigma)(\sigma - i\bar\sigma)),
\ee
where an irrelevant overall constant has been suppressed. At finite $N$
the pre-exponential factors can be expanded as a binomial series which provides
us with an expansion in powers of $m$. The coefficients are elementary
integrals, and for the $m-$dependent part of the partition function we find
\be
\frac {Z(m)}{Z(0)} = 1 + \frac {m^2N^2 \Sigma^2}{4}
+ \frac {m^4 N^4\Sigma^4}{64}(1-\frac 2N) + \cdots.
\ee
It should be noted that no approximations have been made. On the other hand,
the fermion determinant can be written as a product over the eigenvalues
which leads to the expansion
\be
\frac {Z(m)}{Z(0)} = 1 + m^2 \left<\sum_n\frac 1{\lambda^2_n}\right >_{\nu=0}
 + m^4
\frac 12 \left <\sum_{n\ne n'} \frac 1{\lambda_n^2\lambda_{n'}^2}\right >_{\nu
= 0} + \cdots,
\ee
where the average is with respect to the massless partition function.
By equating the coefficients of the powers of $m^2$ we obtain sum rules
for the inverse powers of the eigenvalues that are valid for
any value of $N$. The sum rule (3.3) for $N_f =1$ is  reproduced.
It is valid for any value of $N$. The second sum rule
is modified by the factor $(1-2/N)$. For $N=2$ we find zero which is
correct because in this case there are no terms that contribute to
the sum $n'\ne n$ in (A.3).
\vfill
\eject
\newpage

\vglue 0.6cm
{\bf \noindent  Acknowledgements \hfil}
\vglue 0.4cm
 The reported work was partially supported by the US DOE grant
DE-FG-88ER40388 and by KBN grant PB267512.
We acknowledge the NERSC at Lawrence Livermore where
most of the computations presented in this paper were performed.
Sections 4, 5 and 6 were published in part in \cite{SHURYAK-VERBAARSCHOT-1993}
and \cite{VERBAARSCHOT-ZAHED-1993} for which I
acknowledge my co-workers E.~Shuryak and I. Zahed. I thank
the organizers of the 1993 Zakopane summer school for their hospitality.
Particular thanks goes to A. Bialas and M.A. Nowak. M.A. Nowak is also thanked
for useful discussions and a critical reading of the manuscript.
Finally, I would like to thank A. Smilga for useful discussions.
\vfill
\eject
\newpage
\addtocounter{page}{2}
\noindent
{\bf Figure Captions}
\vskip 0.5cm
\noindent
Fig. 1. The eigenvalue density $n(\lambda)$ for $N_f =0,\,1,\,2,\,$ and 3.
The area below each curve is normalized to 1, and the binsize is 0.00125
for $N_f = 0$ and 0.0025 in the other cases.
\vskip 0.5cm
\noindent
Fig. 2. The microscopic limit of the spectral density in Fig. 1. We show the
density for the
unfolded spectrum where the {\it average} position of the $n$'th eigenvalue
is at $\mu =n$. The dashed line represents
 the exact microscopic spectral density
$\rho_S(\mu)$ for the same normalization, $i.e.$ $\Sigma = \pi$.

\vfill
\eject
\newpage
\addtocounter{page}{-3}
\setlength{\baselineskip}{15pt}

\bibliographystyle{aip}

\end{document}